\documentclass[sigconf]{acmart}
\AtBeginDocument{%
  \providecommand\BibTeX{{%
    \normalfont B\kern-0.5em{\scshape i\kern-0.25em b}\kern-0.8em\TeX}}}

\copyrightyear{2023}
\acmYear{2023}
\setcopyright{acmlicensed}\acmConference[CIKM '23]{Proceedings of the 32nd ACM International Conference on Information and Knowledge Management}{October 21--25, 2023}{Birmingham, United Kingdom}
\acmBooktitle{Proceedings of the 32nd ACM International Conference on Information and Knowledge Management (CIKM '23), October 21--25, 2023, Birmingham, United Kingdom}
\acmPrice{15.00}
\acmDOI{10.1145/xxxxxxx.xxxxxxx}
\acmISBN{979-8-4007-0124-5/23/10}

\begin{document}

\title{STGIN: Spatial-Temporal Graph Interaction Network for Large-scale POI Recommendation}

\author{Shaohua Liu}
\authornote{These authors contributed equally to this research.}
\orcid{0009-0000-4933-1802}
\affiliation{%
  \institution{Meituan}
  \city{Shanghai}
  \country{China}
}
\email{liushaohua07@meituan.com}

\author{Yu Qi}
\authornotemark[1]
\orcid{0009-0001-1135-0597}
\affiliation{%
  \institution{Meituan}
  \city{Shanghai}
  \country{China}
}
\email{qiyu07@meituan.com}

\author{Gen Li}
\authornotemark[1]
\orcid{0009-0007-0908-182X}
\affiliation{%
  \institution{Meituan}
  \city{Shanghai}
  \country{China}
}
\email{ligen08@meituan.com}

\author{Mingjian Chen}
\orcid{0009-0004-9365-8684}
\affiliation{%
  \institution{Meituan}
  \city{Shanghai}
  \country{China}
}
\email{chenmingjian@meituan.com}

\author{Teng Zhang}
\orcid{0009-0009-4199-2935}
\affiliation{%
  \institution{Meituan}
  \city{Shanghai}
  \country{China}
}
\email{zhangteng09@meituan.com}

\author{Jia Cheng}
\orcid{0000-0003-1702-4263}
\affiliation{%
  \institution{Meituan}
  \city{Shanghai}
  \country{China}
}
\email{jia.cheng.sh@meituan.com}

\author{Jun Lei}
\orcid{0000-0002-4015-8668}
\affiliation{%
  \institution{Meituan}
  \city{Shanghai}
  \country{China}
}
\email{leijun@meituan.com}

\renewcommand{\shortauthors}{Shaohua Liu et al.}

\begin{abstract}
In Location-Based Services, Point-Of-Interest(POI) recommendation plays a crucial role in both user experience and business opportunities. Graph neural networks have been proven effective in providing personalized POI recommendation services. However, there are still two critical challenges. First, existing graph models attempt to capture users' diversified interests through a unified graph, which limits their ability to express interests in various spatial-temporal contexts. Second, the efficiency limitations of graph construction and graph sampling in large-scale systems make it difficult to adapt quickly to new real-time interests. To tackle the above challenges, we propose a novel Spatial-Temporal Graph Interaction Network. Specifically, we construct subgraphs of spatial, temporal, spatial-temporal, and global views respectively to precisely characterize the user's interests in various contexts. In addition, we design an industry-friendly framework to track the user's latest interests. Extensive experiments on the real-world dataset show that our method outperforms state-of-the-art models. This work has been successfully deployed in a large e-commerce platform, delivering a 1.1\% CTR and 6.3\% RPM improvement.
\end{abstract}



\maketitle

\section{Introduction}
In recent years, Location-Based Service providers including Facebook, Foursquare, and UberEats have become more and more popular. As one of the key services of LBS providers, the POI recommendation utilizes past user behaviors and contextual POI information to make a personalized recommendation. Unlike the traditional recommendation system, the quality of the POI recommendation is intrinsically linked to three dimensions of data: personal, spatial, and temporal, as well as their mutual interactions\cite{PoiSurvey2022}. For example, a user may order coffee in the morning on business days while looking for a gym on Saturday afternoons.

Many approaches based on sequential user behavior data have been proposed to recommend POI candidates. 
LSTPM\cite{LSTPM2020} explores the temporal and spatial correlations from the long-term behavior sequence and captures geographical influence from short-term sequences. STPIL\cite{STPIL2021} constructs various sequences to acquire the spatial-temporal periodic interests of different granularities, then applies integration for multiple interests. CatDM\cite{CatDM2020} divides a user's check-in history into several time windows and applies a personalized attention mechanism for each time window. However, these works only take into account the user's own behaviors which could cause data sparsity issues in certain spatial-temporal contexts.

Inspired by the idea of collaborative filtering\cite{NGCF2019} that similar users tend to make similar choices, Graph Neural Network\cite{GCN2017,Graphsage2017} has proven to be an effective way to mitigate data sparsity. GE\cite{GE2016} and JLGE\cite{JLGE2018} jointly learn the embeddings of multiple bipartite graphs into the same latent space. STGCN\cite{STGCN2020} considers both user-region periodic pattern and user-POI periodic pattern and fuses all the context information into a unified graph. STPUDGAT\cite{STP-UDGAT2020} leverages spatial, temporal, and preference factors from both local and global views to learn POI-POI relationships. While these methods take advantage of graph structure to improve representation learning, they fail to explicitly model the user's changing interests across different spatial-temporal contexts. Moreover, due to the heavy cost of graph construction and sampling in large-scale industrial systems, these methods have great difficulty in integrating graph representation with the user's real-time behaviors, which can lead to performance degradation in online services.

Motivated by the above analysis, we propose a novel Spatial-Temporal Graph Interaction Network for large-scale POI recommendation. Specifically, a subgraph representation framework has been proposed to learn the user's interests from the spatial view, temporal view, spatial-temporal view, and global view respectively. In addition, to capture the user's latest interests in a real-time manner, we devise a flexible mechanism to combine the interests of multiple views with real-time behaviors in an industrial-friendly way. Our main contributions can be summarized as follows:

\begin{itemize}
\item We propose a spatial-temporal graph interaction method to capture the user's diverse interests under different spatial-temporal contexts. To the best of our knowledge, it is the first work to use multi-view spatial-temporal subgraphs for POI recommendation.
\item We propose an industrial-friendly framework that combines the spatial-temporal graph learning and the user's real-time behaviors in an end-to-end manner and is able to track the user's latest preferences.
\item We have successfully deployed this work in a large location-based e-commerce platform and achieved encouraging results in both online and offline experiments.
\end{itemize}

\section{Problem definition}
Generally speaking, a recommender system consists of two stages: matching and ranking\cite{YoutubeDNN2016}. In this paper, we implement our method in the matching stage by learning the embedding representations of queries and POIs respectively. With very few modifications, our model is also available in the ranking stage.

Let $\mathcal{U}$ denote the set of users, $\mathcal{P}$ denote the set of candidate POIs, $\mathcal{S}$ denote the set of locations, and $\mathcal{T}$ denote the set of time slots that are divided by a specific pattern(like an hour, day, etc).  ${p^u} = \{p_1^u, p_2^u, ..., p_{L}^u\}$ denotes the behavior sequence of the user $u$. It is a list of POIs that are ordered by the corresponding behavior timestamps. Here $L$ denotes the length of the user $u$’s behavior sequence. In this work, we consider clicking as the behavior type.

Since users are consistently interacting with new POIs, we divide the behavior sequence $p^u$ into two parts, one is the real-time sequence ($p_r^u$) which grows with new click behaviors, the other behaviors are considered as the history sequence ($p_h^u$).

For a query $q = (u, s, t)$ requested by a user $u$ under current location $s$ and time $t$, the goal of the POI recommendation is to select the top $K$ POIs ($\mathcal{P}'$) that the user would be interested in. It can be formulated as
\begin{equation}
\mathop{\arg\max}_{\mathcal{P}' \subset \mathcal{P}, |\mathcal{P}'|=K}\sum_{p \in \mathcal{P}'} sim(\boldsymbol{e_q}, \boldsymbol{e_p}),
\end{equation}
where $\boldsymbol{e_q}$ is the vector representation of the query $q$, and $\boldsymbol{e_p}$ is the vector representation of the POI $p$. The function $sim(·)$ calculates the similarity of two vectors.

\begin{figure*}[h]
  \centering
  \vspace{-12pt}
  \setlength{\abovecaptionskip}{+3pt}
  \setlength{\belowcaptionskip}{-5pt}
  \includegraphics[width=\textwidth]{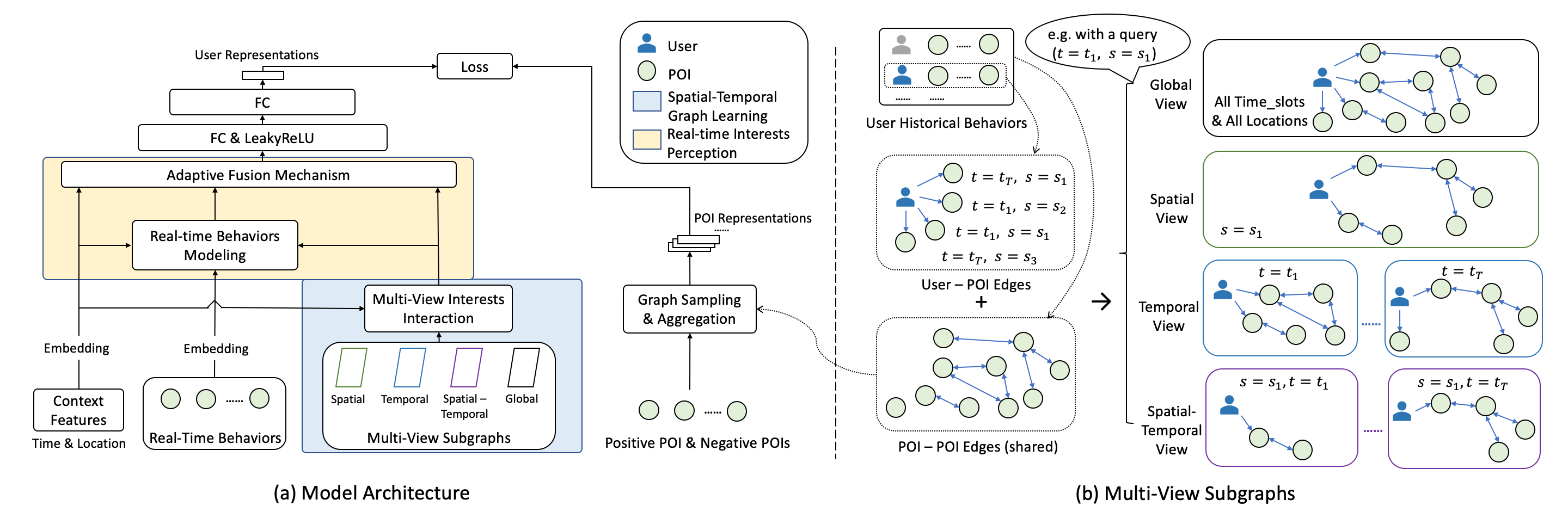}
  \caption{Overall network structure and construction of multi-view subgraphs. }
  \Description{Model Structure}
  \label{Fig.1}
\end{figure*}

\section{Methodology} 
In this section, we elaborate on the proposed Spatial-Temporal Graph Interaction Network. Model structure is shown in Figure~\ref{Fig.1}(a).

\subsection{Spatial-Temporal Graph Learning}
Users tend to show different preferences in different spatial-temporal contexts. Since a unified graph mixes information from various contexts, it is difficult to accurately express the user's interest in a specific spatial-temporal context. Therefore, we use multi-view subgraphs for context-specific interest learning.\\
\textbf{Graph Construction.} Based on the user's history sequence $p_h^u$, we construct subgraphs from the four types of views: global view, spatial view, temporal view, and spatial-temporal view. Each subgraph has two types of nodes, representing users and POIs in our task, connected by two types of heterogeneous edges. \textit{\textbf{Clicking edges}} represent users' explicit interests. As shown in Figure~\ref{Fig.1}(b), we extract POIs from behavior sequences that occurred in the corresponding spatial-temporal contexts and build an edge between the user and each POI. Note that, clicking edges in the global view are constructed with POIs in all contexts. \textit{\textbf{Co-clicking edges}} represent the implicit relation between POIs. An edge is constructed between two POIs that have been clicked by the same user within a session\cite{DSIN2019}. It should be emphasized that co-clicking edges are shared among all subgraphs. \\
\textbf{Graph Sampling\,\&\,Aggregation.} To precisely capture the user's interest in each view, we devise a spatial-temporal oriented mechanism for node aggregation. Specifically, for each user $u \in \mathcal{U}$, we adopt meta-path\cite{metapath2vec2017} ``user $\xrightarrow[]{click}$ POI $\xrightarrow[]{co-click}$ POI'' to generate neighbors. Such a sampling strategy not only preserves the uniqueness of each view but also enhances the representation through shared co-clicking edges. Taking the temporal view $t_i \in \mathcal{T}$ as an example, neighbor set $B_{t_i}^u$ contains both POIs clicked by the user $u$ at time $t_i$ and POIs co-clicked by others within a short period. 
Then we apply a two-layer GAT\cite{GAT2018} mechanism to get the user's interest in this temporal view as follows
\begin{equation}
\boldsymbol{u_{t_i}} = GAT\_AGG(\{p | p \in B_{t_i}^u\}).
\end{equation}
Similarly, we can get the user's interest from the spatial view $\boldsymbol{u_{s_j}}$, the spatial-temporal view $\boldsymbol{u_{s_j t_i}}$, and the global view $\boldsymbol{u_g}$.
The neighbors for each POI are generated from the meta-path ``POI $\xrightarrow[]{co-click}$ POI $\xrightarrow[]{co-click}$ POI'', note that we skip the user node when generating POI's neighbors, since a user may have different types of behaviors in the sequence, which may introduce noise for learning the POI. Again, we use a two-layer GAT mechanism to get the representation of the POI as $\boldsymbol{e_p}$, which is shared across all views.\\ 
\textbf{Multi-view Interests Interaction.}
Different contexts play different roles in determining each user's interests. Therefore, we design a user-dependent attention module to learn the relations among different views. For all spatial views, since POIs located far away are hardly reachable, we fetch the user's spatial view that exactly matches the current location. For all temporal views, we apply an attention aggregation process to learn their mutual influence guided by the user's global view. The process is defined as
\begin{equation}
\boldsymbol{u_t} = \sum_{i=1}^{T}\alpha_{t_i}\cdot \boldsymbol{u_{t_i}}.
\end{equation}
Note that $\alpha_{t_i}$ is the attention weight and is defined as
\begin{equation}
\alpha_{t_i} =
\frac{exp(f(\boldsymbol{W_t^T}[\boldsymbol{e}_{|t_i - t_q|} || \boldsymbol{u_{t_i}} ||
\boldsymbol{u_g}]))}{\sum_{k=1}^{T}exp(f(\boldsymbol{W_t^T}[\boldsymbol{e}_{|t_k - t_q|} || \boldsymbol{u_{t_k}} || \boldsymbol{u_g}]))},
\end{equation}
where $f(.)$ is the activation function, || indicates concatenation, $\boldsymbol{e}_{|t_i-t_q|}$ denotes the embedding of the absolute difference between query time $t_q$ and the time of the corresponding temporal view $t_i$, $T$ denotes the number of time slots, $\boldsymbol{W_t^T}$ is the weighting matrix.

The same process can be applied for spatial-temporal views and get $\boldsymbol{u_{st}}$ as the desired output.

Finally, we get the user's multi-view interest through all views
\begin{equation}
\boldsymbol{u_h} =f(\boldsymbol{W_h^T}[\boldsymbol{u_g}||\boldsymbol{u_s}||\boldsymbol{u_t}||\boldsymbol{u_{st}}]).
\end{equation}

\subsection{Real-time Interests Perception}
Users are constantly interacting with new POIs, and these real-time behaviors are essential for predicting users' subsequent interests. Therefore, we need to develop a flexible real-time interest tracking mechanism compatible with spatial-temporal graph learning.\\
\textbf{Real-time Behaviors Modeling.}
Users' real-time behaviors ($p_r^u$) indicate their latest needs, but historical habits and current context should not be overlooked. First, historical habits such as brands and lifestyles often play an important role in current decisions. Second, given the context of the current time and location, users' next visit is likely to be related to a subset of real-time behaviors. 
Consequently, we model the user's real-time interest evolution based on the above factors
\begin{equation}
\boldsymbol{u_r} = \sum_{i=1}^{M}\alpha_{r_i}\cdot \boldsymbol{e_{p_{r_i}}}.
\end{equation}
$\boldsymbol{e_{p_{r_i}}}$ denotes the representation of a recently clicked POI. $M$ is the length of real-time behaviors. $\alpha_{r_i}$ is the attention weight, which is formalized as
\begin{equation}
\alpha_{r_i} = \frac{exp(f(\boldsymbol{W_r^T}[\boldsymbol{e}_{|t_{r_i} - t_q|}||\boldsymbol{e}_{|s_{r_i} - s_q|}||\boldsymbol{e_{p_{r_i}}}||\boldsymbol{u_h}]))}{\sum_{k=1}^{M}exp(f(\boldsymbol{W_r^T}[\boldsymbol{e}_{|t_{r_k} - t_q|}||\boldsymbol{e}_{|s_{r_k} - s_q|}||\boldsymbol{e_{p_{r_k}}}|| \boldsymbol{u_h}]))},
\end{equation}
where || indicates concatenation, $\boldsymbol{e}_{|s_{r_i} - s_q|}$ denotes the embedding of the distance between query location $s_q$ and the location of clicked POI $s_{r_i}$, $\boldsymbol{W_r^T}$ is the weighting matrix.\\
\textbf{Adaptive Fusion Mechanism.}
we conduct a gating-based fusion to generate the final query representation $\boldsymbol{e_q}$
\begin{equation}
\boldsymbol{e_q} = \alpha_{1} \cdot \boldsymbol{u_h} + \alpha_{2} \cdot \boldsymbol{u_r} + \alpha_{3} \cdot [\boldsymbol{e_{t_q}}||\boldsymbol{e_{s_q}}].
\end{equation}
$\alpha_{1}$, $\alpha_{2}$, $\alpha_{3}$ are generated by the following softmax function
\begin{equation}
[\alpha_{1}, \alpha_{2},\alpha_{3}] = Softmax([\boldsymbol{w_h^T}\boldsymbol{u_h},\boldsymbol{w_r^T}\boldsymbol{u_r},\boldsymbol{w_c^T}[\boldsymbol{e_{t_q}}||\boldsymbol{e_{s_q}}]]),
\end{equation}
where $\boldsymbol{w_h^T}$, $\boldsymbol{w_r^T}$, and $\boldsymbol{w_c^T}$ are weighting vectors, $\boldsymbol{e_{t_q}}$ and $\boldsymbol{e_{s_q}}$ are the embeddings of query time $t_q$ and query location $s_q$.

\subsection{Optimization}
After obtaining the representations $\boldsymbol{e_q}$ and $\boldsymbol{e_p}$
, we adopt the pairwise training method\cite{RankNet2005} and minimize the hinge loss as follows
\begin{equation}
Loss = \sum_{i=1}^{N} \sum_{j=1}^{K} max(0,margin - sim(\boldsymbol{e}_q^{i},\boldsymbol{e}_{p_i}^{+})+sim(\boldsymbol{e}_q^{i},\boldsymbol{e}_{p_j}^{-})),
\end{equation}
where $N$ is the length of the training data and $K$ is the number of negative samples for each positive sample.

\section{Implementation}\label{Implementation}
\textbf{Training}. Graph building and learning are the most time-consuming parts in \verb|STGIN|. Though Cartesian combinations between spatial and temporal information may be numerous, the actual volume of graph data is acceptable since most users only visited a small number of locations. Thus the graph can be trained simultaneously with other parts. In practice, we divide timestamps into four slots (morning, noon, dinnertime, and night) and use geohash\footnote{https://en.wikipedia.org/wiki/Geohash} of length 5 to divide a city into multiple locations. For our production dataset with over 300 million users, 20 million POIs, and 600 million clicking behaviors, it takes about four hours to train on 50 virtual machines, each providing 6 CPU cores and 40GB RAM.\\
\textbf{Serving}. Benefiting from the real-time interests perception module, the graph structure could be updated at a given frequency without real-time requirements. The learned embeddings are stored in a key-value table, thus we can skip the real-time inference of graphs, which may become a bottleneck in online service. It's worth noting that having a new clicking POI will change a user's real-time sequence as well as the representation vector $\boldsymbol{e_q}$. On average, it takes just 5.6 milliseconds to make an online recommendation.

\section{Experiments}
We design experiments to answer the following research questions: 
(RQ1): How does STGIN perform compared to other state-of-the-art models? (RQ2): What are the effects of different components? (RQ3): How effective are the spatial-temporal views interacting with each other? (RQ4): How does STGIN perform in online A/B tests? 

\subsection{Experimental Setup}
\begin{table}
  \setlength{\abovecaptionskip}{+5pt}
  \setlength{\belowcaptionskip}{-7pt}
  \caption{Statistics of dataset}
  \label{tab:statistics}
  \begin{tabular}{cccccc}
    \toprule
    Data & Users & POIs & Records & BehaviorLen & Locations\\
    \midrule
    Train & 637563 & 108551 & 2023304 & 149.6 & 320\\
    \hline
    Test & 217561 & 60982 & 532194 & 149.9 & 305\\
  \bottomrule
\end{tabular}
\end{table}

\begin{table}
  \setlength{\abovecaptionskip}{+5pt}
  \setlength{\belowcaptionskip}{-5pt}
  \caption{Comparisons of different methods}
  \label{tab:exp_result}
  \begin{tabular}{lll}
    \toprule
    Methods&HitRate@200&Recall@200\\
    \midrule
    {\verb|CatDM|} & 21.37\% & 15.42\%\\
    {\verb|STPIL|} & 23.80\% & 17.45\%\\
    {\verb|JLGE|} & 19.96\% & 14.45\% \\
    {\verb|STGCN|} & 18.52\% & 13.14\%\\
    {\verb|STPUDGAT|} & 22.21\% & 16.04\% \\
    \hline
    {\verb|STGIN|} & $\boldsymbol{25.87}\%$ & $\boldsymbol{19.12}\%$ \\
    {\verb|STGIN-RT|} & 19.71\% & 14.10\% \\
    {\verb|STGIN-Temporal|} & 22.62\% & 16.35\% \\
    {\verb|STGIN-Spatial|} & 25.12\% & 18.57\% \\
    {\verb|STGIN-Interaction|} & 22.56\% & 16.26\% \\
  \bottomrule
\end{tabular}
\end{table}
\textbf{Dataset and Settings}. We conduct offline experiments on industrial production data of a large e-commerce platform. It spans one month(2022-08-09 to 2022-09-08) and is sampled in a few cities. The statistics of dataset are shown in Table~\ref{tab:statistics}. We take behaviors over the last 24 hours as the real-time sequence.
All experiments are trained on Tensorflow with graph learning engine Euler\footnote{https://github.com/alibaba/euler}. We adopt Adam\cite{Adam2015} as the optimizer and the dimension of each trainable feature embedding is 16. The batch size and learning rate are set to 1024 and 0.001. We apply LeakyReLU\cite{LeakyReLU2010} as the activation function. The number of negative samples is 6.\\
\textbf{Competitors}. We implement several competitive baselines for evaluation. {\verb|JLGE|}\cite{JLGE2018} expresses user, POI, location, and time in the same latent space through multiple bipartite graphs. {\verb|STGCN|}\cite{STGCN2020} fuses all spatial-temporal information into a unified graph and applies a time-based neighborhood sampling algorithm. {\verb|STPUDGAT|}\cite{STP-UDGAT2020} uses both spatial and temporal factors to model the POI-POI relationship. Besides, we also implement two popular sequential models. 
{\verb|CatDM|}\cite{CatDM2020} and {\verb|STPIL|}\cite{STPIL2021} both explore user interests from spatial-temporal behavior subsequences and integrate them through attention mechanisms. To further understand the contribution of each component, we design four variants of our model. {\verb|STGIN-RT|} drops real-time behavior sequence. {\verb|STGIN-Temporal|} drops subgraphs of the temporal view. {\verb|STGIN-Spatial|} drops subgraphs of the spatial view. {\verb|STGIN-Interaction|} merely concatenates graph representation and real-time sequence without their mutual interaction.\\
\textbf{Metrics}. We apply widely used \textbf{HitRate@K} and \textbf{Recall@K}\cite{EBR2020,SDM2019,GRU4REC2015} to evaluate the performance of all methods.\\

\vspace{-12pt}
\subsection{Offline Evaluation}
\begin{figure}[h]
  \centering
  \vspace{-10pt}
  \setlength{\abovecaptionskip}{+5pt}
  \setlength{\belowcaptionskip}{-10pt}
  \includegraphics[width=\linewidth]{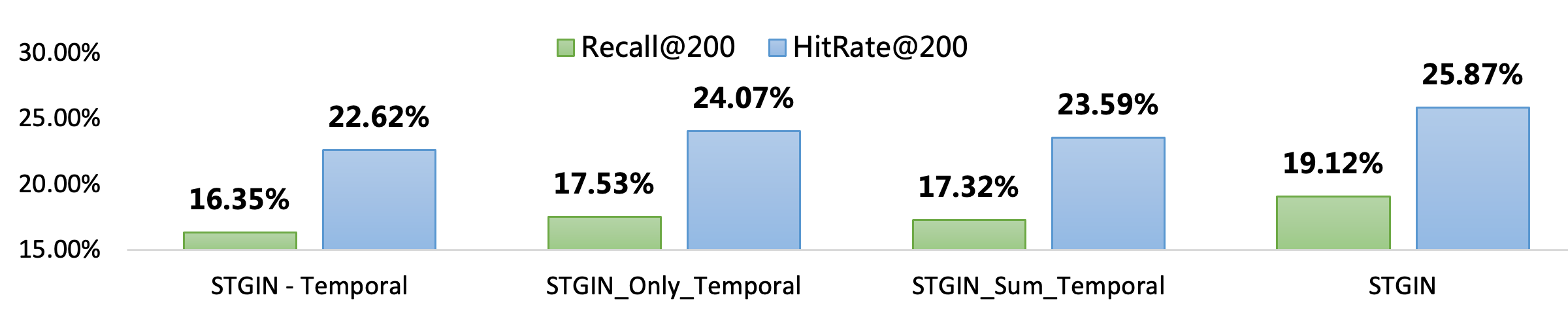}
  \caption{Comparisons of temporal-related subgraphs.}
  \label{fig:teporal_subgraph}
\end{figure}

\textbf{Exp-RQ1}. As shown in Table~\ref{tab:exp_result}, {\verb|STGIN|} outperforms other competitors in all metrics. Due to the heavy cost of graph construction and sampling, {\verb|STGCN|} faces a major challenge in addressing users' real-time behaviors. {\verb|JLGE|} and {\verb|STPUDGAT|} try to solve this challenge with the POI-POI graph but pay little attention to the interest interaction.
In addition, a unified graph cannot express fine-grained interests in different spatial-temporal contexts.
 {\verb|CatDM|} and {\verb|STPIL|} are good at dealing with users' real-time behaviors, but they cannot address the issues of data sparsity when users have limited behaviors.\\
\textbf{Exp-RQ2}. From table~\ref{tab:exp_result}, all the components significantly benefit our task. Precisely, {\verb|STGIN-RT|} demonstrates real-time behaviors are essential for POI recommendation. {\verb|STGIN-Temporal|} and {\verb|STGIN-Spatial|} show that it is useful to capture the interests of users from spatial-temporal views. {\verb|STGIN-Interaction|} shows that the potential of real-time behaviors can be enhanced by interaction.\\
\textbf{Exp-RQ3}. To validate the effects of interaction among spatial-temporal views, we design temporal-related experiments from coarse to fine-grained. {\verb|STGIN_Only_Temporal|} only retrieves a temporal subgraph that matches the current time. {\verb|STGIN_Sum_Temporal|} retrieves temporal subgraphs of all time slots and applies sum pooling operation. In Figure~\ref{fig:teporal_subgraph}, {\verb|STGIN-Temporal|} gets the lowest score because it omits temporal information. {\verb|STGIN_Only_Temporal|} performs slightly better than {\verb|STGIN_Sum_Temporal|}, for the latter method does not differentiate the user's attention on different temporal views. Consequently, {\verb|STGIN|} achieves the best performance.

\begin{table}
  \setlength{\abovecaptionskip}{+5pt}
  \setlength{\belowcaptionskip}{-7pt}
  \caption{Result for Online A/B Test}
  \label{tab:online_result}
  \begin{tabular}{ccl}
    \toprule
    Metric&CTR&RPM\\
    \midrule
    Relative  Improvement & +1.1\% & +6.3\%\\
  \bottomrule
\end{tabular}
\vspace{-7pt}
\end{table}

\vspace{-3pt}
\subsection{Online A/B Test (RQ4)}
Based on the optimizations described in the section \ref{Implementation}, we have successfully deployed the \verb|STGIN| model in our production environment to handle the real traffic of a large e-commerce APP. Online evaluation metrics are CTR and RPM (Revenue Per Mille). As noted in Table~\ref{tab:online_result}, both metrics increased compared to the base model, demonstrating the effectiveness of 
\verb|STGIN| in practical environments.

\vspace{-3pt}
\section{Conclusion}
In this paper, we propose a novel approach for POI recommendation called \verb|STGIN|. It can characterize a user's diverse interests from spatial-temporal subgraphs and capture a user's latest interests through real-time behaviors modeling. Both online and offline experiments demonstrate the effectiveness of our method.

\bibliographystyle{ACM-Reference-Format}
\balance
\bibliography{reference}

\end{document}